\documentclass[reprint,,longbibliography,preprintnumbers,superscriptaddress,prd,
nofootinbib,amsmath,amssymb,aps]{revtex4-1}

\usepackage{graphicx}
\usepackage{dcolumn}
\usepackage{bm}
\usepackage{float}
\usepackage{overpic}
\usepackage[english]{babel}
\usepackage{graphicx}
\usepackage{epsfig}
\usepackage{amssymb}
\usepackage{amsmath}
\usepackage[plainpages=false,pagebackref=false,pdftex]{hyperref}
\usepackage{ulem}
\usepackage[T1]{fontenc}
\usepackage[utf8]{inputenc}
\usepackage{color}
\usepackage[nameinlink,capitalize]{cleveref}

\usepackage[titletoc]{appendix}
\usepackage{subfig}

\usepackage{booktabs}

\usepackage{comment}

\begin{document}

\title{Thermal properties of the $\rho$ meson from a machine-learning-improved holographic QCD model}

\author{Jia-Jie Jiang}
\author{Han Tang}
\author{Danning Li}
\email{Corresponding author: lidanning@jnu.edu.cn}
\vspace{1cm}
\affiliation{Department of Physics and Siyuan Laboratory, Jinan University, Guangzhou 510632, China}

\author{Lang Yu}
\email{Co-corresponding author: yulang@jlu.edu.cn}
\affiliation{College of Physics, Jilin University, Changchun 130012, P.R. China}

\begin{abstract}
The thermal properties of the \(\rho\) meson are revisited within the soft-wall AdS/QCD framework. The dilaton background is calibrated against lattice QCD screening-mass data through a neural-network-assisted inverse procedure. Using this calibrated background, we compute the temporal vector spectral function and the lowest vector quasi-normal modes. The pole mass \(m_{\mathrm{pole}}(T)\) and thermal width \(\Gamma(T)\) are then extracted from the corresponding complex quasi-normal mode frequencies. 
Adopting the vanishing real part of the quasi-normal mode frequency, 
\(\omega_{\mathrm{Re}}(T_d)=0\), as a criterion for meson dissociation, we estimate 
\(T_d\approx168\,\mathrm{MeV}\). As the temperature increases, the thermal width grows, and the spectral function evolves from a narrow Breit-Wigner-like resonance peak into a transport peak, confirming the QNM prediction that the vanishing of the pole mass is accompanied by the emergence of a transport peak and making it difficult to identify a quasiparticle peak. A numerical extraction of the dispersion relation shows that the propagator can be described by a finite-temperature simple-pole form with momentum-dependent thermal width, characterized by a diffusion coefficient \(D(T)\) and an effective propagation speed \(u(T)\), whose temperature dependences are also extracted. This form modifies the usual relation between the screening and pole masses. These results may provide a useful reference for effective studies of the hot hadronic medium.
\end{abstract}

\keywords{$\rho$ meson, thermal spectral properties, holographic QCD, machine learning}

\maketitle

\section{Introduction}\label{introduction}

The thermal properties of hadrons are of essential importance, as hadrons serve as crucial probes of the hot and dense QCD medium created in relativistic heavy-ion collisions. Understanding these properties is vital for unraveling the behavior of matter under extreme conditions, such as those encountered in the early universe or in heavy-ion collision experiments. A particularly sensitive example is the pion, the pseudo-Goldstone boson associated with chiral symmetry breaking, whose in-medium pole-mass modifications provide valuable insights into chiral symmetry restoration ~\cite{Pisarski:1983ms,Son:2001ff}; low-momentum pion excesses observed in heavy-ion collisions may also receive contributions from charged-pion condensation ~\cite{ALICE:2013mez,Begun:2015ifa}. These observations motivate detailed studies of the thermal properties of hadrons.

Complementary to the pseudoscalar channel, vector mesons provide a different probe of the hot and dense medium through their in-medium spectral functions, which govern the electromagnetic response of the medium ~\cite{Rapp:2007bm,Geurts:2022xmk}. The lightest vector meson, the \(\rho\) meson, is therefore of particular interest. Its in-medium spectral function dominates the low-mass dilepton spectrum, so modifications of its properties---such as a broadening of the peak or a shift in its peak position---can provide valuable information about the hot medium and its possible connection to chiral symmetry restoration~\cite{Rapp:1999ej,vanHees:2007th}.

Lattice QCD studies of the vector channel associated with the \(\rho\) meson indicate a consistent qualitative temperature trend ~\cite{Karsch:2003jg,Cheng:2010fe,Ding:2012ar,Bazavov:2019www,Krasniqi:2024kwm}. At temperatures below the chiral crossover temperature \(T_c\), the vector screening mass is nearly independent of temperature and remains close to the zero-temperature \(\rho\)-meson pole mass, while above \(T_c\) it rises rapidly with temperature. In the spectral function, the \(\rho\)-meson peak broadens strongly and loses spectral weight as the temperature rises. At high temperatures, fits to lattice correlators indicate that a low-frequency transport peak begins to emerge, with spectral weight migrating toward low frequencies, consistent with the dissolution of the \(\rho\) meson ~\cite{Brandt:2015aqk}. The ill-posed nature of analytic continuation leaves the quantitative thermal width and peak shape uncertain, and the detailed structure of the low-frequency region depends on the spectral reconstruction method. Despite these uncertainties, the overall temperature dependence---from a stable bound state at low \(T\), through the transition region near \(T_c\), to broadening and dissociation above \(T_c\)---provides an important qualitative benchmark for effective models. This motivates the use of effective models calibrated by lattice inputs.

Beyond lattice QCD, the Nambu-Jona-Lasinio (NJL) framework has also been used to investigate the thermal properties of the $\rho$ meson~\cite{He:1997gn,Liu:2016vuw,Liu:2015pna,Liu:2014uwa}. Approaches based on Dyson-Schwinger and Bethe-Salpeter equations (DSE/BSE) and on the functional renormalization group (FRG) have provided complementary pictures across temperature regimes~\cite{Gao:2020hwo,Chen:2020afc,Tripolt:2021jtp,Jung:2016yxl}. QCD sum rules and hadronic many-body theory also offer tests and predictions for the in-medium properties of the $\rho$ meson~\cite{Kwon:2010fw,Ayala:2012ch,Urban:1999im,Eletsky:2001bb}. These calculations generally show that the $\rho$-meson spectral function broadens with increasing temperature~\cite{Chen:2020afc,Jung:2016yxl,Eletsky:2001bb}, and that its line shape deviates from a narrow Breit-Wigner resonance, with spectral weight shifting toward low frequencies~\cite{Jung:2016yxl}. However, the operational definitions of dissociation are not uniform across different frameworks. Various approaches employ different criteria to identify the onset of melting, leading to disparate values of $T_d$ and no commonly accepted benchmark~\cite{Ayala:2012ch,Rapp:2009yu}. 
Predictions for the in-medium $\rho$-meson mass also exhibit different temperature dependences: nearly constant~\cite{Mallik:1997kj}, increasing~\cite{Gao:2020hwo,Harada:1996pg,Hofmann:1999nn}, or decreasing with temperature~\cite{He:1997gn,Ayala:2013vra,Zhang:2016qrl}.
These differences reflect the model dependence of the predictions and the different definitions of the in-medium meson mass, motivating further investigations.

Holographic methods are particularly suited to studying real-time dynamics in strongly coupled systems~\cite{Son:2007vk}. The quasi-normal modes~\cite{Berti:2009kk} of the bulk black-hole geometry correspond one-to-one to the poles of the retarded Green's function~\cite{Son:2002sd} in the dual field theory, directly characterizing relaxation, transport, and the disappearance of spectral peaks~\cite{Kovtun:2005ev}. The high-precision numerical determination of quasi-normal modes is also supported by mature algorithms such as the continued fraction method~\cite{Daghigh:2022uws}. These features make holography well suited for studying the thermal $\rho$ meson. Within the soft-wall AdS/QCD framework~\cite{Karch:2006pv}, which offers several advantages---it reproduces approximately linear Regge trajectories in vacuum~\cite{Karch:2006pv,Batell:2008zm,Li:2013oda,Vega:2008te,Ballon-Bayona:2024yuz,Forkel:2007cm}, is computationally efficient, and naturally incorporates external parameters such as temperature, density, and (isospin) chemical potential for systematic investigations~\cite{Li:2013oda,Batell:2008zm,Bartz:2016ufc,Cao:2020ryx,Liang:2023lgs,Zhao:2022uxc,Fang:2018axm}---the thermal properties of light mesons have been described with reasonable consistency~\cite{Cao:2021tcr,Ballon-Bayona:2024twa,Cao:2020ryx}. 
The spectral functions at finite temperature typically exhibit temperature-dependent shifts in their peak positions, accompanied by peak broadening~\cite{Fujita:2009wc,Cao:2021tcr,Mamani:2013ssa}.
Direct calculations of quasi-normal modes in the vector channel reproduce the evolution from narrow peaks at low temperatures, to broader structures at intermediate temperatures, and finally to the disappearance of quasiparticle peaks at high temperatures~\cite{Cao:2021tcr,Mamani:2013ssa}. 
Cross-channel comparative studies reveal different thermal width behaviors in the vector and axial-vector channels~\cite{Cao:2021tcr}, while the scalar and pseudoscalar channels exhibit related thermal behavior near the chiral transition~\cite{Cao:2021tcr,Cao:2020ryx}.

However, in these soft-wall studies, the dilaton field background enters as an input of the construction. As a result, the predicted thermal properties, such as the meson dissociation behavior ~\cite{Fujita:2009wc,Mamani:2013ssa,Bartz:2016ufc,Cao:2021tcr,Giannuzzi:2012ux} and the temperature dependence of screening masses~\cite{Cao:2021tcr}, are tied to the chosen dilaton profile. A common strategy is to first fix the model parameters by fitting the vacuum hadron spectrum~\cite{Li:2013oda,Rinaldi:2021dxh} and then extrapolate to finite temperature~\cite{Cao:2021tcr,Cao:2020ryx}. A more desirable strategy may be to constrain the dilaton background using lattice data, so that the finite-temperature behavior is anchored by first-principles input rather than by a vacuum-based extrapolation. This is where machine learning can play a complementary role. Given its broad applications to nuclear-physics inverse problems~\cite{Boehnlein:2021eym,Zhou:2023pti}, machine learning allows physical models or underlying parameters to be reconstructed from limited observables~\cite{Carleo:2019ptp}. Neural ordinary differential equations (Neural ODEs)~\cite{chen2018neural} have further expanded the scope of equation solving and inverse modeling, especially for complex physical systems. For example, machine learning has been applied to improve holographic models using experimental meson spectra~\cite{Akutagawa:2020yeo,ThomasArun:2025uyi} and lattice QCD and related thermodynamic data~\cite{Chen:2024ckb,Zhu:2026dmc,Cai:2024eqa}.

Building on the soft-wall framework and recent machine-learning techniques~\cite{RAISSI2019686,chen2018neural}, we enhance the soft-wall model through a data-driven approach: lattice screening masses $m_{\rm scr}(T)$~\cite{Bazavov:2019www} are used to determine the dilaton field through a neural-network approach~\cite{LeCun:2015pmr}. 
Using this unified background, we compute the complex quasi-normal mode frequencies in the vector channel, from which the temperature-dependent pole mass $m_{\rm pole}(T)$ and thermal width $\Gamma(T)$ are extracted.
Adopting the vanishing real part of the quasi-normal mode frequency, $m_{\mathrm{pole}}=0$, as an operational dissociation criterion, we estimate the $\rho$ meson dissociation temperature to be $T_d\approx168\,\mathrm{MeV}$.
Through an analysis of the dispersion relation and the retarded propagator, we adopt the finite-temperature parameterization introduced in Ref.~\cite{Chen:2025ncu},
\[
\omega(p)= -i\,\bigl[a+D\,\vec{p}^{\,2}\bigr] \ \pm\ \sqrt{u^{2}\vec{p}^{\,2}+m^{2}},
\]
and determine its parameters from our numerical dispersion relation. 
Here \(a\) characterizes the momentum-independent damping, \(D\) is the diffusion coefficient, \(u\) is the effective propagation speed, and \(m^2\) is a parameter of the dispersion relation.
This yields a medium-modified relation between \(m_{\rm scr}\) and \(m_{\rm pole}\), extending previous formulations~\cite{Mamani:2013ssa,Fayazbakhsh:2012vr,Fayazbakhsh:2013cha,Sheng:2020hge}. The temperature dependences of \(D\) and \(u\) are also extracted. These results provide a useful reference for effective studies of the hot hadronic medium, and the same data-driven framework can be extended to finite density or isospin chemical potential in future work.

The rest of this paper is organized as follows:~\autoref{sec:softwall} introduces the finite-temperature soft-wall AdS/QCD framework and our machine-learning pipeline described in \autoref{subsec:ml}.
Subsequently,~\autoref{sec:thermal} builds on the calibrated dilaton field to compute the vector-channel quasi-normal modes (\autoref{subsec:qnm}), spectral functions (\autoref{subsec:spectral}), and dispersion relation (\autoref{subsec:dispersion}).
Finally,~\autoref{sec:conclusion} summarizes the main findings of this work and outlines possible directions for future research.
The appendices provide the vector-field equations of motion, discuss the longitudinal screening mass associated with the temporal component \(V_0\), and present the coupled \(V_0\)-\(V_3\) equations governing the longitudinal channel for general \((\omega,\vec p)\).

\section{Soft-wall AdS/QCD model at finite temperature}\label{sec:softwall}

To compute the thermal properties of the $\rho$ meson, we adopt the finite-temperature soft-wall AdS/QCD framework and calibrate its dilaton background using lattice QCD data. In this section we first recall the model setup and the background geometry, and then describe the machine-learning pipeline used to determine the dilaton field.

We work in the bottom-up soft-wall AdS/QCD framework~\cite{Karch:2006pv}, in which the five-dimensional (5D) bulk theory is dual to a four-dimensional (4D) strongly coupled gauge theory with $\mathrm{SU}_L(2)\times\mathrm{SU}_R(2)$ chiral symmetry. In this approach, the Regge-like spectrum of light mesons is reproduced by introducing a dilaton field background $\Phi(z)$ that breaks the conformal invariance in the infrared, while the chiral gauge fields $L_M$ and $R_M$ encode the dynamics of vector and axial-vector mesons. The scalar field $X$, which is dual to the chiral condensate, is responsible for spontaneous chiral symmetry breaking.

The background geometry is a five-dimensional asymptotically AdS black-brane spacetime,
\begin{equation}
\begin{aligned}
ds^2
&= e^{2A(z)}\!\Big[f(z)\,dt^2 - dx^{i}dx_{i} - \tfrac{dz^2}{f(z)}\Big],
\end{aligned}
\label{eq:bluk}
\end{equation}
where $z$ is the holographic radial coordinate, with the boundary located at $z=0$ and the horizon at $z=z_h$. The warp factor and blackening function are
\begin{equation}
A(z)=-\ln (z),
\qquad
f(z)=1-\frac{z^{4}}{z_{h}^{4}},
\end{equation}
respectively. The Hawking temperature of the black brane is determined by the surface gravity at the horizon,
\begin{equation}
T = \left| \frac{f'(z_h)}{4\pi} \right| = \frac{1}{\pi z_h} .
\label{eq:Tzh}
\end{equation}

The dilaton field is usually taken as \(\Phi(z)=\mu_g^2 z^2\), where \(\mu_g\) is a phenomenological parameter fixed by the light-meson spectrum. In this work, however, we calibrate \(\Phi(z)\) directly from lattice QCD screening-mass data, as described in \autoref{subsec:ml}.

With 5D indices $M,N=0,\ldots,4$ and $\sqrt{g}$ the metric determinant, the action of the model takes the form
\begin{equation}
\begin{gathered}
S = \int d^5x \sqrt{g}\,e^{-\Phi(z)} \,\mathrm{Tr}\bigl[
|D_M X|^2 - V(|X|)\\
- \frac{1}{4g_5^2}\left(F^2_{L} + F^2_{R}\right)\bigr],
\label{eq:action}
\end{gathered}
\end{equation}
where the gauge coupling constant is taken as $g_5=2\pi$ for $N_c=3$~\cite{Erlich:2005qh}. The matrix-valued scalar field $X$ transforms under $\mathrm{SU}_L(2)\times\mathrm{SU}_R(2)$, and its covariant derivative is
\begin{equation}
D_M X = \partial_M X - i L_M X + i X R_M .
\end{equation}
The chiral gauge fields are expanded as $L_M=L_M^i t^i$ and $R_M=R_M^i t^i$ with generators $t^i=\sigma^i/2$ $(i=1,2,3)$, and their field strengths read
\begin{equation}
\begin{aligned}
F^L_{MN} = \partial_M L_N - \partial_N L_M - i [L_M,L_N], \\
F^R_{MN} = \partial_M R_N - \partial_N R_M - i [R_M,R_N] .
\end{aligned}
\end{equation}
It is convenient to decompose the chiral gauge fields into vector and axial-vector fields,
\begin{equation}
V_M^i = \tfrac{1}{2}\left(L_M^i + R_M^i\right),
\qquad
A_M^i = \tfrac{1}{2}\left(L_M^i - R_M^i\right),
\label{eq:VA}
\end{equation}
in terms of which the field strengths become
\begin{equation}
\begin{aligned}
F^V_{MN} = \partial_M V_N - \partial_N V_M - i [V_M,V_N] - i [A_M,A_N],\\
F^A_{MN} = \partial_M A_N - \partial_N A_M - i [V_M,A_N] - i [A_M,V_N],
\label{eq:FVA}
\end{aligned}
\end{equation}
and the covariant derivative of $X$ becomes
\begin{equation}
D_M X = \partial_M X - i [V_M,X] - i \{A_M,X\}.
\label{eq:DMX}
\end{equation}
The scalar potential is
\begin{equation}
V(|X|) = m_5^2 |X|^2 + \lambda |X|^4 ,
\end{equation}
with $m_5^2$ the 5D mass and $\lambda$ a free parameter. Taking the degenerate up and down quark masses, $m_q = m_u = m_d$, the scalar field takes the form
\begin{equation}
X = \frac{\chi}{2}\, I .
\label{eq:Xchi}
\end{equation}

To study fluctuations, we decompose the 4D vector fluctuation into a divergence-free part and a gradient part,
\begin{equation}
\begin{gathered}
v^i_\mu = v^{T,i}_\mu + \partial_\mu \varphi^i,\qquad \\
\partial^\mu v^{T,i}_\mu = 0,
\end{gathered}
\end{equation}
and fix the radial gauge $v_z = 0$. For the static screening problem, choosing the spatial momentum along the $x^3$ direction, we focus on the transverse spatial components $v_1^i$ and $v_2^i$. The quadratic action reads
\begin{align}
S_v
&= -\frac{1}{2g_5^2}\!\int d^5x\,\sqrt{g}\,e^{-\Phi}
\sum_{i=1}^3 \Big\{
g^{zz}g^{\mu\nu}\,\partial_z v^i_\mu\,\partial_z v^i_\nu
\notag\\[-2pt]
&\qquad\qquad\quad
+\, g^{\mu\nu}g^{mn}\,\partial_\mu v^i_m\,\partial_\nu v^i_n
\Big\}.
\label{eq:Sv}
\end{align}
Starting from this quadratic action, we linearize the non-Abelian field strength and keep only the vector sector. Using the AdS$_5$ metric Eq.~\eqref{eq:bluk}, the transverse spatial mode $v(z)$ in the static limit $\omega=0$ obeys
\begin{equation}
v'' + \left( A' + \frac{f'}{f} - \Phi' \right) v'
      - \frac{p^{2}}{f}\, v = 0 .
 \label{eq:v-ODE}
\end{equation}
Here we only consider the transverse mode, and the full analysis will be left in the appendix.
The asymptotic behavior near the UV boundary can be obtained as
\begin{equation}
v(z \to 0) \;=\;
  v_{0} \;+\; v_{2}\, z^{2}
  \;+\; \frac{1}{2}\, v_{0} p^{2} z^{2}\log(z)
  \;+\; \mathcal{O}(z^{3}) ,
  \label{eq:v-UV}
\end{equation}
while the near-horizon expansion yields
\begin{equation}
v(z \to z_{h}) \;=\;
  v_{h0} \;+\; \frac{p^{2} z_{h}}{4}\, v_{h0}\,(z_{h}-z)
  \;+\; \mathcal{O}\!\bigl[(z_{h}-z)^{2}\bigr] .
  \label{eq:v-IR}
\end{equation}
Here \(v_0\) is the source, \(v_2\) is the response coefficient, and \(v_{h0}\) is the horizon value of the regular solution; the overall normalization of \(v\) is arbitrary because the equation is linear and homogeneous, and it can be set as \(v_{h0}=1\).

The screening mass $m_{\rm scr}$ is obtained from the solution of Eq.~\eqref{eq:v-ODE} that is regular at the horizon. In the long-distance limit, this solution governs the exponential decay of the spatial correlation function, $G(x)\sim e^{-m_{\rm scr}|x|}$. Given this setup, the main adjustable component is the dilaton field $\Phi(z)$, and we will employ machine learning techniques to calibrate $\Phi(z)$ so as to achieve quantitative agreement with lattice QCD data. We note that the other parameters in the soft-wall model do not affect the vector spectral function, which is why we start from this channel.

\subsection{Machine-learning Calibration}\label{subsec:ml}

In this subsection, we calibrate the dilaton field \(\Phi(z)\) using a neural network (NN)~\cite{LeCun:2015pmr}. The input is the lattice screening mass \(m_{\mathrm{scr}}(T)\), whose relation to the bulk geometry can be summarized as follows. The screening mass is defined through the pole of the static spatial correlator, \(m_{\mathrm{scr}}^{2}=-\,p_v^{2}\), where \(p_v\) is obtained by solving the equation of motion Eq.~\eqref{eq:v-ODE} subject to the UV boundary condition Eq.~\eqref{eq:v-UV} and the near-horizon regularity condition Eq.~\eqref{eq:v-IR}. Since the horizon location \(z_h\) is fixed by the temperature through Eq.~\eqref{eq:Tzh}, each pair \((T,m_{\mathrm{scr}})\) defines a boundary-value problem for the dilaton-dependent equation of motion. Numerically, we integrate Eq.~\eqref{eq:v-ODE} from the IR toward the UV using the regular near-horizon expansion and require that the resulting solution satisfies the UV condition Eq.~\eqref{eq:v-UV}; this procedure is repeated for all temperatures in the input set. For later convenience, we introduce \(k\equiv p_v^{2}\), so that the pole condition reads \(k=-\,m_{\mathrm{scr}}^{2}\).

\begin{figure}[H]
  \centering
  \includegraphics[width=0.5\textwidth]{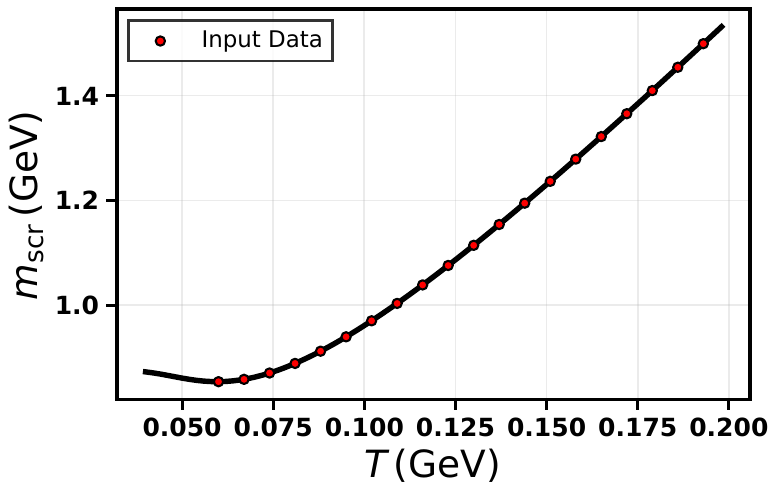}
  \caption{Test data for validating the machine-learning procedure. The black solid line shows the screening mass computed with the known dilaton field $\Phi(z) = \mu_g^2 z^2$ ($\mu_g = 0.44\,\text{GeV}$), replicated from Ref.~\cite{Cao:2021tcr}, with the red points marking the 20 selected input points. The corresponding numerical values are listed in Table~\ref{tab:0-input-data-test-table}.}
  \label{fig:0-input-data-test}
\end{figure}

\begin{table}[H]
  \centering
  \begingroup
  \setlength{\tabcolsep}{6pt}
  \begin{ruledtabular}
  \begin{tabular}{cccc}
    $T$ (GeV) & $m_{\mathrm{scr}}$ (GeV) & $T$ (GeV) & $m_{\mathrm{scr}}$ (GeV) \\
    \hline
    0.0600 & 0.8541 & 0.1300 & 1.1143 \\
    0.0670 & 0.8585 & 0.1370 & 1.1540 \\
    0.0740 & 0.8706 & 0.1440 & 1.1948 \\
    0.0810 & 0.8889 & 0.1510 & 1.2364 \\
    0.0880 & 0.9123 & 0.1580 & 1.2788 \\
    0.0950 & 0.9397 & 0.1650 & 1.3219 \\
    0.1020 & 0.9703 & 0.1720 & 1.3655 \\
    0.1090 & 1.0034 & 0.1790 & 1.4096 \\
    0.1160 & 1.0388 & 0.1860 & 1.4542 \\
    0.1230 & 1.0758 & 0.1930 & 1.4991 \\
  \end{tabular}
  \end{ruledtabular}
  \caption{The numerical values of the selected test data for validating the machine-learning procedure.}
  \label{tab:0-input-data-test-table}
  \endgroup
\end{table}

\begin{figure}[H]
  \centering
  \includegraphics[width=0.5\textwidth]{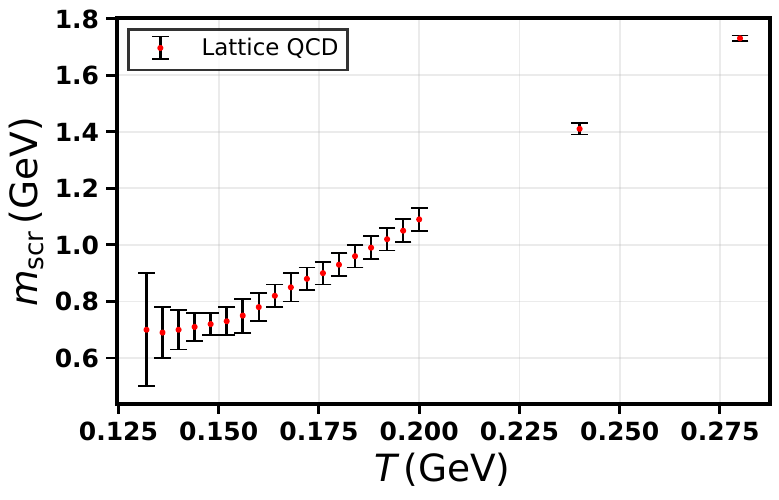}
  \caption{Lattice QCD data~\cite{Bazavov:2019www} used as the actual training input for the neural network. The numerical values are listed in Table~\ref{tab:0-input-data-LQCD-table}.}
  \label{fig:0-input-data-LQCD}
\end{figure}

\begin{table}[H]
  \centering
  \begingroup
  \setlength{\tabcolsep}{6pt}
  \begin{ruledtabular}
  \begin{tabular}{cccc}
    $T$ (GeV) & $m_{\mathrm{scr}}$ (GeV) & $T$ (GeV) & $m_{\mathrm{scr}}$ (GeV) \\
    \hline
    0.1320 & 0.7000 & 0.1720 & 0.8800 \\
    0.1360 & 0.6900 & 0.1760 & 0.9000 \\
    0.1400 & 0.7000 & 0.1800 & 0.9300 \\
    0.1440 & 0.7100 & 0.1840 & 0.9600 \\
    0.1480 & 0.7200 & 0.1880 & 0.9900 \\
    0.1520 & 0.7300 & 0.1920 & 1.0200 \\
    0.1560 & 0.7500 & 0.1960 & 1.0500 \\
    0.1600 & 0.7800 & 0.2000 & 1.0900 \\
    0.1640 & 0.8200 & 0.2400 & 1.4100 \\
    0.1680 & 0.8500 & 0.2800 & 1.7300 \\
  \end{tabular}
  \end{ruledtabular}
  \caption{Lattice QCD data for the screening mass $m_{\mathrm{scr}}(T)$ from Ref.~\cite{Bazavov:2019www}, used as the actual training input for the neural network. }
  \label{tab:0-input-data-LQCD-table}
  \endgroup
\end{table}

We first show the input data used in this work (Figures~\ref{fig:0-input-data-test} and~\ref{fig:0-input-data-LQCD}); the numerical values are listed in Tables~\ref{tab:0-input-data-test-table} and~\ref{tab:0-input-data-LQCD-table}.
In Figure~\ref{fig:0-input-data-test}, the black solid line shows the results replicated from Ref.~\cite{Cao:2021tcr}, which are used as test data to validate our machine learning model.
From this curve, we select 20 points (marked in red) as inputs for the inversion; their numerical values are listed in Table~\ref{tab:0-input-data-test-table}.
Figure~\ref{fig:0-input-data-LQCD} shows the lattice QCD data~\cite{Bazavov:2019www}; the numerical values are listed in Table~\ref{tab:0-input-data-LQCD-table}. We use the central values of the lattice QCD data as fixed inputs and do not propagate the quoted uncertainties. A full uncertainty propagation is left to future work. Specifically, we take as inputs the first 20 lattice points with $T\le 0.28~\mathrm{GeV}$ reported in Table~X (the $m_V$ column) of Ref.~\cite{Bazavov:2019www}.

To ensure the validity of our machine learning model, we first obtain test data by computing the dilaton field $\Phi(z) = \mu_g^2 z^2$ (with $\mu_g=0.44\,\text{GeV}$), and input this test data $(T,m_{\mathrm{scr}})$ to verify whether the program can recover the same dilaton field. Subsequently, keeping all hyperparameters fixed, we replace the input data with the lattice QCD data $(T, m_{\mathrm{scr}})$ to obtain the physically meaningful dilaton field. In this way, the dilaton field is determined solely by the input data.

We use a neural network to represent the dilaton field $\Phi(z)$. The model solves the vector-mode ODE system for all temperatures in a batch over $\{T_i\}$. It jointly learns $\Phi(z)$ and a parameter $k_i$ for each temperature. We employ a high-accuracy ODE integrator and a two-stage optimizer: an Adam warm-up followed by L-BFGS. The training loss has four components: a soft UV regularization on the dilaton field, a UV boundary penalty for the vector mode, a smoothing penalty on $\Phi''(z)$, and a data-consistency term. The data-consistency term softly regularizes $k_i$ toward $-m_i^2$. We initialize $k_i=-m_i^2$ but do not fix it as a hard constraint during training.

The dilaton field is parametrized by a multilayer perceptron. The network has three hidden layers with 64 units each and \texttt{tanh} activations, followed by a linear output layer. This architecture is denoted as \texttt{[1,64,64,64,1]}, where the first and last entries are the input and output dimensions. To enforce the expected near-boundary behavior, the UV conditions in Eq.~\eqref{eq:uv-soft-phi} are imposed. In addition, the smoothing regularizer in Eq.~\eqref{eq:smooth} is included to suppress nonphysical oscillations in $\Phi(z)$.
These constraints ensure that the neural network output $\Phi_{\mathrm{NN}}(z)$ is physically reasonable while maintaining sufficient model flexibility.

For each temperature $T_i$, we set $z_{h,i}=1/(\pi T_i)$ from Eq.~\eqref{eq:Tzh} and introduce a normalized radial coordinate
\begin{equation}
\begin{gathered}
x \equiv \frac{z}{z_{h,i}} ,\\
\frac{\mathrm{d}}{\mathrm{d}x} = z_{h,i}\,\frac{\mathrm{d}}{\mathrm{d}z}.
\label{eq:x-def}
\end{gathered}
\end{equation}
After this rescaling, the integration is performed over $x\in[1-\Delta,\Delta]$ with $\Delta=10^{-8}$, i.e., from the IR to the UV. To handle all temperatures efficiently, the ODE solving is carried out simultaneously for all $T_i$ in a batched manner.

With the normalized coordinate, we rewrite the second-order equation for $v(z)$, Eq.~\eqref{eq:v-ODE}, as the first-order system
\begin{equation}
\begin{gathered}
y=\begin{pmatrix} v \\ v_z \end{pmatrix},\\
\frac{\mathrm{d}y}{\mathrm{d}x}= z_h \begin{pmatrix} v_z \\ v_{zz} \end{pmatrix},\\
v_{zz}= -\Bigl(A'(z)-\Phi'(z)+\frac{f'(z)}{f(z)}\Bigr)v_z + \frac{k}{f(z)}\,v.
\end{gathered}
\label{eq:fos}
\end{equation}
Here, $k$ denotes the trainable parameter $k_i$ at temperature $T_i$. Although we initialize $k_i$ at $-m_i^2$ for numerical convenience, it is not fixed during training; instead, a quadratic penalty $(m_i^2 + k_i)^2$ softly regularizes $k_i$ toward $-m_i^2$.

Since the equation is linear and homogeneous in $v$, we fix the overall normalization by setting $v_{h0}=1$. At the IR endpoint $x_{\mathrm{IR}}=1-\Delta$ (with $z_{\mathrm{IR}}=x_{\mathrm{IR}}\,z_h$), we use the IR regular expansion
\begin{equation}
\begin{gathered}
v(z_{\mathrm{IR}})=1+\frac{k z_h}{4}\,(z_h-z_{\mathrm{IR}}),
\qquad\\
v'(z_{\mathrm{IR}})=-\frac{k z_h}{4}.
\label{eq:IRfirst}
\end{gathered}
\end{equation}
At the UV endpoint $x_{\mathrm{UV}}=\Delta$ (with $z_{\mathrm{UV}}=x_{\mathrm{UV}}\,z_h$), a soft penalty on $v(z_{\mathrm{UV}})^2$ approximates the condition $v(0)=0$. In addition, near the UV boundary ($z\to 0$), we impose a soft regularization on the dilaton,
\begin{equation}
\Phi(0) = 0,\qquad \Phi'(0) = 0.
\label{eq:uv-soft-phi}
\end{equation}

To improve identifiability and numerical stability, for each temperature $T_i$ we uniformly sample $N_i=300$ points in $z\in\bigl[10^{-3},\,(1-10^{-3})\,z_{h,i}\bigr]$ and include the smoothing loss
\begin{equation}
\begin{gathered}
\mathcal{L}_{\mathrm{smooth}}
= \lambda_{\mathrm{smooth}}\,
\frac{1}{\sum_i N_i}\sum_{i=1}^{N_{\rm temp}}\sum_{j=1}^{N_i}
\bigl[\Phi''(z_{i,j})\bigr]^2,\\
\quad \lambda_{\mathrm{smooth}}=10^{-3}.
\label{eq:smooth}
\end{gathered}
\end{equation}
Here $N_{\rm temp}$ denotes the number of temperature measurements, i.e., the number of input data points.

For solving the ODE, we do not prescribe a fixed integration mesh; since $\Phi_{\mathrm{NN}}(z)$ is continuous and the solver is adaptive (controlled by \texttt{rtol} and \texttt{atol}), the evaluation nodes along $z$ are chosen automatically. For visualization and error inspection only, we additionally evaluate $\Phi_{\mathrm{NN}}(z)$ on $N_{\mathrm{plot}}=600$ uniformly spaced points over the regular radial interval.

The total loss is given by
\begin{equation}
\begin{gathered}
\mathcal{L}
= \lambda_{\mathrm{BC}}\bigl[\Phi(0)^2+\Phi'(0)^2\bigr]
+ \sum_i \Bigl\{\, v_i(x_{\mathrm{UV}})^2  \Bigr\}\\
+ \sum_i \Bigl\{\bigl(m_i^2+k_i\bigr)^2 \Bigr\}
+ \mathcal{L}_{\mathrm{smooth}},\\
\qquad \lambda_{\mathrm{BC}}=10.0.
\label{eq:loss}
\end{gathered}
\end{equation}

For the numerical integration, we employ batched ODE solving with \texttt{torchdiffeq.odeint}, using the DOPRI5 (Dormand-Prince) method~\cite{DORMAND198019,Paszke:2019xhz}. We adopt a two-stage tolerance schedule: Stage~1 with $(\texttt{rtol},\texttt{atol})=(10^{-6},10^{-8})$, and Stage~2 with $(\texttt{rtol},\texttt{atol})=(10^{-10},10^{-12})$. Optimization proceeds with Adam (500 steps, learning rate $10^{-3}$) for warm-up, followed by L-BFGS with a maximum of 500 iterations, using a strong Wolfe line search and tolerances \texttt{tolerance\_grad} $=10^{-12}$ and \texttt{tolerance\_change} $=10^{-12}$.

\begin{figure}[H]
  \centering
  \includegraphics[width=0.5\textwidth]{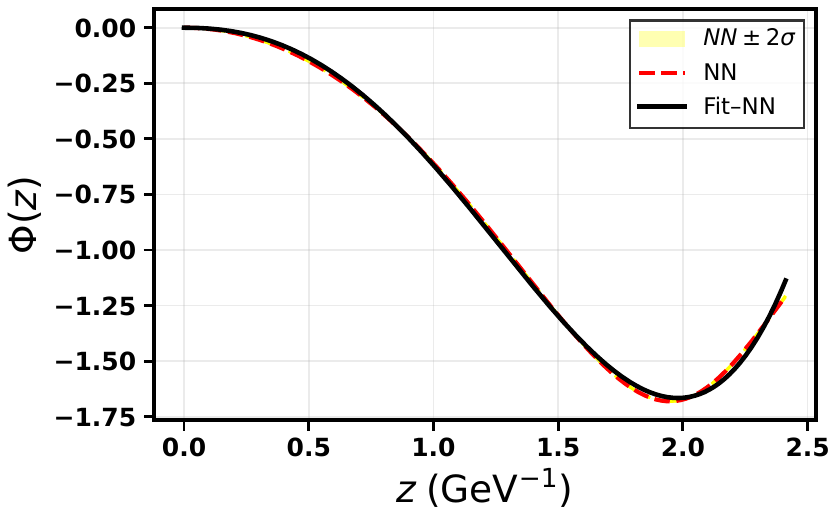}
  \caption{The dilaton field profile averaged over 10 runs (red dashed line), obtained from the neural network using the lattice QCD data~\cite{Bazavov:2019www} listed in Table~\ref{tab:0-input-data-LQCD-table} as input. The yellow band shows the mean $\pm 2\sigma$ across the 10 runs. For later convenience, we fit the neural-network output with a polynomial (black solid line), yielding
$\Phi(z) = -0.3473\,z^{2} - 0.5079\,z^{3} + 0.2367\,z^{4}$,
which we denote by ``Fit-NN'' in what follows.\protect\footnotemark[1]}
  \label{fig:2-NN-LQCD-fig}
\end{figure}

On the test dataset, we perform a single training run and carry out a basic consistency check by reproducing the target $\Phi(z) = \mu_g^2 z^2$ with $\mu_g=0.44\,\text{GeV}$. For the lattice QCD dataset, we conduct 10 independent runs with different random seeds and report the run-averaged $\Phi(z)$. The fixed base seed is \texttt{BASE\_SEED}$=20250802$. Per-run child seeds are generated with \texttt{numpy.random.SeedSequence} in the batch launcher, and each seed is passed to the training script via \texttt{--seed}. Inside the script, the same seed initializes the RNGs of NumPy, Python's \texttt{random}, and PyTorch (CPU/CUDA). We show both the 10-run mean and the mean $\pm 2\sigma$ band in Figure~\ref{fig:2-NN-LQCD-fig}.

\begin{figure}[H]
  \centering
  \includegraphics[width=0.5\textwidth]{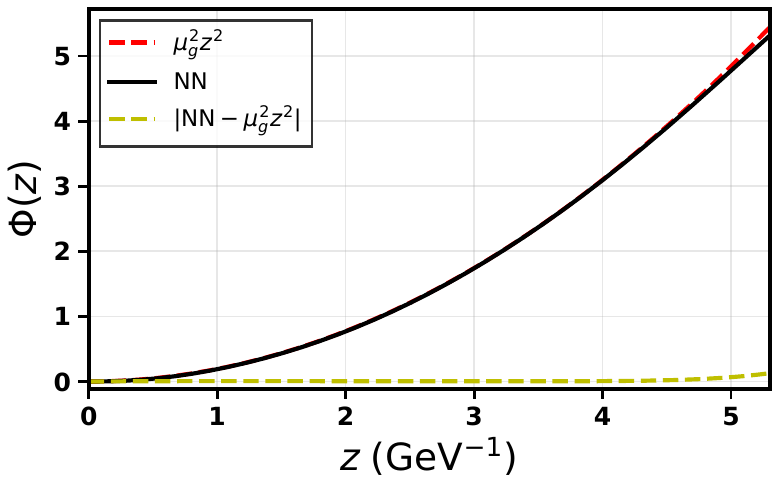}
  \caption{Comparison of the neural network result (black solid line) from the test data in Table~\ref{tab:0-input-data-test-table} with the true solution $\Phi(z) = \mu_g^2 z^2$ ($\mu_g=0.44\,\text{GeV}$) (red dashed line). The yellow dashed line shows the difference $\bigl|\Phi_{\mathrm{NN}}(z)-\Phi(z)\bigr|$.}
  \label{fig:2-NN-Test-fig}
\end{figure}

\footnotetext[1]{Independent repetitions of the complete 10-run training procedure may lead to small batch-to-batch variations in the last digits of the fitted coefficients of $\Phi(z)$ due to the numerical neural-network optimization and ODE integration. The coefficients quoted here correspond to the 10-run batch used throughout the subsequent analysis.}

\begin{figure}[H]
  \centering
  \includegraphics[width=0.5\textwidth]{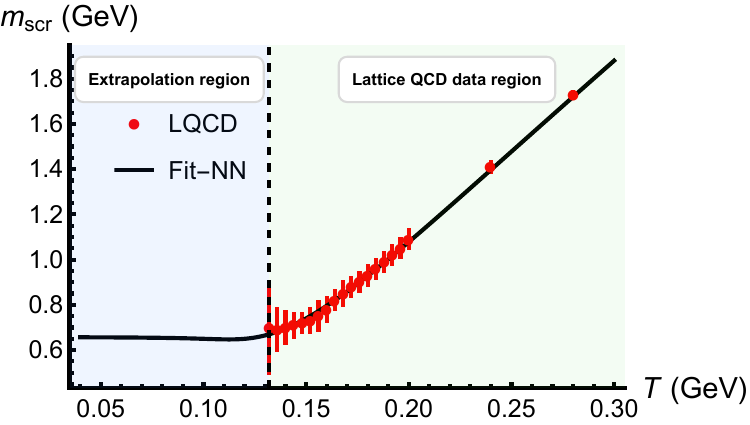}
  \caption{Screening mass from Fit-NN (black solid line) compared with the lattice QCD results (red points with error bars). The shaded regions indicate the extrapolation region (blue) and the lattice QCD data region (green).}
  \label{fig:2-NNmscr}
\end{figure}

We first use the test data generated from $\Phi(z) = \mu_g^2 z^2$ ($\mu_g=0.44\,\text{GeV}$), as listed in Table~\ref{tab:0-input-data-test-table}, to verify the validity of the procedure. The results of the inversion test are shown in Figure~\ref{fig:2-NN-Test-fig}. The red dashed line represents $\Phi(z) = \mu_g^2 z^2$ ($\mu_g=0.44\,\text{GeV}$). The black solid line denotes the neural network reconstruction using the data in Table~\ref{tab:0-input-data-test-table} as input. The yellow dashed line indicates the absolute error between the two (evaluated at 600 points). The figure demonstrates that the neural network method accurately reconstructs the dilaton field.

Having validated the inversion procedure on the test data, we now apply it to the lattice QCD data listed in Table~\ref{tab:0-input-data-LQCD-table}. In Figure~\ref{fig:2-NN-LQCD-fig}, the red dashed line denotes the mean of 10 trainings, and the yellow band shows the mean \(\pm 2\sigma\) over the 10 runs. For convenient subsequent use, we approximate the neural-network output obtained from the lattice QCD data by a quartic polynomial,
\begin{equation}
\Phi_{\mathrm{Fit\text{-}NN}}(z)
= -0.3473\,z^{2} - 0.5079\,z^{3} + 0.2367\,z^{4},
\label{eq:PhiFitNN}
\end{equation}
which we denote by ``Fit-NN'' in what follows.
Using this fitted dilaton field, we recompute the screening masses (see Figure~\ref{fig:2-NNmscr}).
In the temperature region covered by the training data, the model accurately reproduces the inputs.
In the low-temperature extrapolation region, the computed screening mass remains almost unchanged with temperature, consistent with the behavior obtained in other theoretical calculations~\cite{Gao:2020hwo,Krasniqi:2024kwm}. Here, for the low-temperature extrapolation, the computed screening mass is \(m_{\mathrm{scr}}=656~\mathrm{MeV}\) at \(T=40~\mathrm{MeV}\).

\section{Thermal properties of the $\rho$ meson in the calibrated model}\label{sec:thermal}

Using the screening-mass data of the \(\rho\) meson and the machine-learning calibration, we determine the dilaton field of the soft-wall AdS/QCD model in a data-driven manner, achieving quantitative agreement with lattice QCD over the temperature range covered by the data. On this calibrated background, we will perform a unified analysis of the \(\rho\)-meson thermal properties: the pole mass and thermal width will be extracted from the quasi-normal modes in \autoref{subsec:qnm}, the spectral function will be obtained from the retarded correlator in \autoref{subsec:spectral}, and the dispersion relation will be analyzed to determine the finite-temperature propagator in \autoref{subsec:dispersion}.

\subsection{Quasi-normal Modes}\label{subsec:qnm}

The complex quasinormal mode (QNM) frequency is written as
\begin{equation}
    \omega = \omega_{\mathrm{Re}} - i\,\omega_{\mathrm{Im}},
    \label{eq:qnm-frequency}
\end{equation}
where $\omega_{\mathrm{Re}}$ and $\omega_{\mathrm{Im}}$ represent the oscillation frequency and the damping rate of the QNM, respectively. Conventionally, the pole mass and thermal width are defined by
\begin{equation}
    m_{\mathrm{pole}} = \omega_{\mathrm{Re}}, \qquad
    \Gamma = 2\omega_{\mathrm{Im}}.
\end{equation}
This is because, near the pole and in the narrow-width limit, the spectral function takes a Breit--Wigner form proportional to
\begin{equation}
    \frac{1}{(\omega-\omega_{\mathrm{Re}})^2+\omega_{\mathrm{Im}}^2}.
\end{equation}
For details, see, e.g., Ref.~\cite{Cao:2021tcr}. When the thermal width becomes comparable to the pole mass, the real part of the QNM frequency may differ significantly from the spectral peak position. However, for later convenience, we keep this sign convention even when the thermal width is large. As the temperature increases and the thermal width grows, it becomes increasingly difficult to identify a well-defined quasiparticle excitation.

Under the same setup and gauge choice as above, the transverse vector fluctuation at zero spatial momentum ($p=0$) obeys
\begin{equation}
v''(z)+\Bigl(A'(z)+\frac{f'(z)}{f(z)}-\Phi'(z)\Bigr)v'(z)+\frac{\omega^2}{f(z)^2}\,v(z)=0.
\label{eq:v-ODEmg}
\end{equation}
Near the UV boundary, the expansion of Eq.~\eqref{eq:v-ODEmg} reads
\begin{equation}
v(z \to 0) = v_0 + v_2 z^2 - \frac{1}{2} v_0 \omega^2 z^2 \log(z) + \mathcal{O}(z^3),
\label{eq:v-ODEmgz0}
\end{equation}
while near the horizon it takes the ingoing form
\begin{equation}
v(z \to z_h) = (z_h - z)^{-\frac{i \omega}{4 \pi T}} \left\{ v_{h0} + v_{h1}(z_h - z) + \mathcal{O}[(z_h - z)^2] \right\},
\label{eq:v-ODEmgzh}
\end{equation}
with
\begin{equation}
v_{h1}=\frac{\bigl(2 - 4 z_h \Phi'(z_h) - 3 i z_h \omega\bigr)\omega}
           {8\bigl(2 i + z_h \omega\bigr)} v_{h0}.
\end{equation}
Since the equation is linear and homogeneous, we fix the overall normalization by setting $v_{h0}=1$.

Using the near-horizon ingoing expansion in Eq.~\eqref{eq:v-ODEmgzh}, we numerically integrate Eq.~\eqref{eq:v-ODEmg} toward the UV boundary. Together with the near-boundary expansion in Eq.~\eqref{eq:v-ODEmgz0}, the source-free condition $v_0(\omega)=0$ determines the quasi-normal mode frequency.

\begin{figure}[H]
  \centering
  \includegraphics[width=0.5\textwidth]{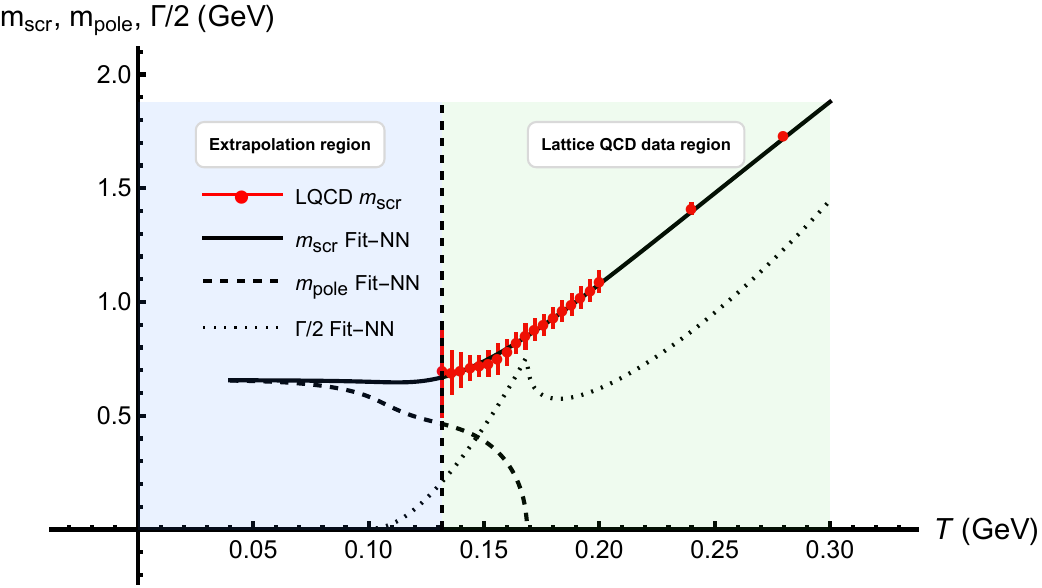}
  \caption{The screening mass (solid curve), pole mass (dashed curve), and half-width $\Gamma/2$ (dotted curve) calculated using the Fit-NN dilaton field $\Phi(z)$, respectively. The red dots with error bars represent the lattice QCD screening mass results from Ref.~\cite{Bazavov:2019www}. }
\label{fig:M-NN-mpg}
\end{figure}

We obtain the dilaton field $\Phi(z)$ and insert it into Eq.~\eqref{eq:v-ODEmg} to compute the vector pole mass $m_{\mathrm{pole}}$ and thermal width $\Gamma$. The results are displayed in Figure~\ref{fig:M-NN-mpg}. The red points with error bars are the lattice QCD results for the screening mass $m_{\mathrm{scr}}$. The black solid, dashed, and dotted curves represent the screening mass, pole mass, and half-width $\Gamma/2$ obtained from the Fit-NN dilaton field $\Phi(z)$, respectively. The light blue and green shaded regions denote the extrapolation region and the lattice QCD data region, respectively. The black solid curve shows $m_{\mathrm{scr}}$, which is nearly flat at low temperatures and agrees with the lattice data at higher temperatures. The black dashed curve shows $m_{\mathrm{pole}}$, which coincides with $m_{\mathrm{scr}}$ below $T\simeq60~\mathrm{MeV}$, starts to decrease at $T\simeq60~\mathrm{MeV}$, reaches zero at $T\simeq168~\mathrm{MeV}$, and remains zero thereafter. The black dotted curve shows $\Gamma/2$, which is small at low temperatures, grows appreciably from $T\simeq102~\mathrm{MeV}$ onward, and exhibits a local maximum near $T\simeq168~\mathrm{MeV}$, followed by a decrease and a subsequent rise.

At low temperatures, we find $m_{\mathrm{scr}}\simeq m_{\mathrm{pole}}$ and a numerically negligible thermal width, indicating narrow bound states. As $T$ increases, $m_{\mathrm{pole}}$ begins to decrease while $m_{\mathrm{scr}}$ remains nearly constant; the width remains small until $T\simeq102~\mathrm{MeV}$ and then grows, signaling in-medium damping. As $T$ approaches $168~\mathrm{MeV}$, the real part of the QNM frequency $\omega_{\mathrm{Re}}$ vanishes, and the width exhibits a pronounced peak. With further heating, $m_{\mathrm{pole}}$ stays zero, the width first decreases and then rises again, and $m_{\mathrm{scr}}$ continues to increase. As shown in Ref.~\cite{Cao:2021tcr}, when $\omega_{\mathrm{Re}}$ vanishes, the low-frequency part of the spectral function is expected to develop a transport peak with the main part $\sim \omega/(\omega^2+\omega_{\mathrm{Im}}^2)$, instead of the Breit-Wigner quasiparticle form $\sim 1/[(\omega-\omega_{\mathrm{Re}})^2+\omega_{\mathrm{Im}}^2]$. This picture is qualitatively consistent with the lattice findings of Ref.~\cite{Brandt:2015aqk}, which indicate the disappearance of a well-defined quasiparticle peak and the dominance of low-frequency spectral weight at high temperature. We therefore adopt the vanishing real part of the quasinormal mode frequency as an operational dissociation criterion, giving a dissociation temperature of $T_d\approx168~\mathrm{MeV}$.

\subsection{Spectral Functions}\label{subsec:spectral}

To cross-check the quasi-normal-mode-based analysis of pole masses and thermal widths, we now compute the spectral function of the transverse vector correlator at zero spatial momentum. The spectral function captures the linear response of the medium to external perturbations, including both resonant features (peaks associated with quasi-normal mode poles) and transport contributions (a low-frequency transport peak), thus offering a more complete thermal picture. Our procedure is as follows: 
we solve the zero-spatial-momentum vector equation in frequency space with ingoing boundary conditions at the horizon, without imposing the source-free condition at the UV boundary; identify $v_0(\omega)$ as the source and $v_2(\omega)$ as the response from the near-boundary expansion; obtain the retarded Green's function $G_v^R(\omega)$; and define the spectral function $\rho(\omega)$ from the imaginary part of $G_v^R(\omega)$.

The near-UV expansion of the bulk vector fluctuation at zero spatial momentum is given in \eqref{eq:v-ODEmgz0}.
After performing holographic renormalization, the retarded Green's function of the transverse vector correlator takes the standard holographic form
\begin{equation}
G_v^R(\omega) = -\frac{1}{g_5^2}\left(\frac{v_2}{v_0}-\frac{\omega^2}{4}\right).
\label{eq:GvR}
\end{equation}
The spectral function, defined from the imaginary part of the retarded Green's function, is thus
\begin{equation}
\rho(\omega) = -\frac{1}{\pi}\operatorname{Im}[G_v^R(\omega)].
\label{eq:spectral}
\end{equation}

For the spectral function, we numerically integrate Eq.~\eqref{eq:v-ODEmg}
from the near-horizon region toward the UV boundary using the ingoing solution
Eq.~\eqref{eq:v-ODEmgzh}.
The numerical solution is matched to the near-boundary expansion
Eq.~\eqref{eq:v-ODEmgz0} by matching both the field and its first derivative,
which determines $v_0(\omega)$ and $v_2(\omega)$.
Substituting these coefficients into Eqs.~\eqref{eq:GvR} and~\eqref{eq:spectral}
yields the spectral function $\rho(\omega)$.

\begin{figure}[H]
\centering
\begin{minipage}{1\textwidth}
  \centering
  \includegraphics[width=\linewidth]{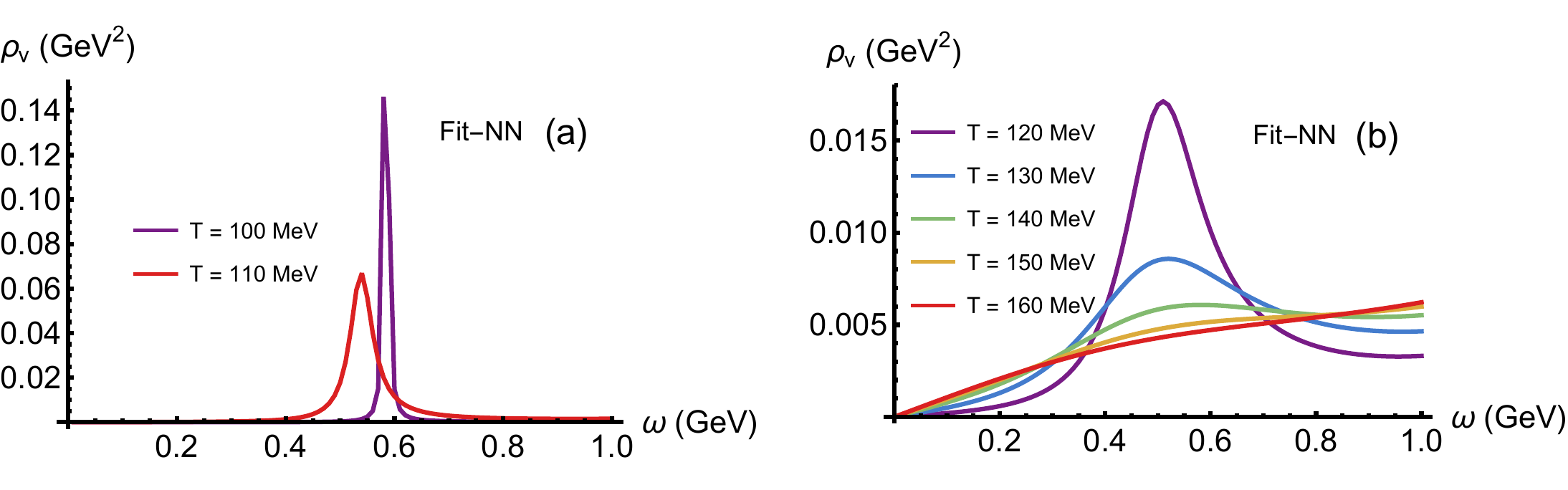}
  \caption{Spectral functions of the transverse vector correlator at several temperatures (\(T=100-160~\mathrm{MeV}\)) computed using the NN-calibrated dilaton field \(\Phi(z)\). All spectral curves are sampled at a uniform frequency spacing of $\Delta\omega=0.01~\mathrm{GeV}$.}
  \label{fig:SFNN}
\end{minipage}
\end{figure}

\begin{figure}[H]
\centering
\begin{minipage}{1\textwidth}
  \centering
  \includegraphics[width=\linewidth]{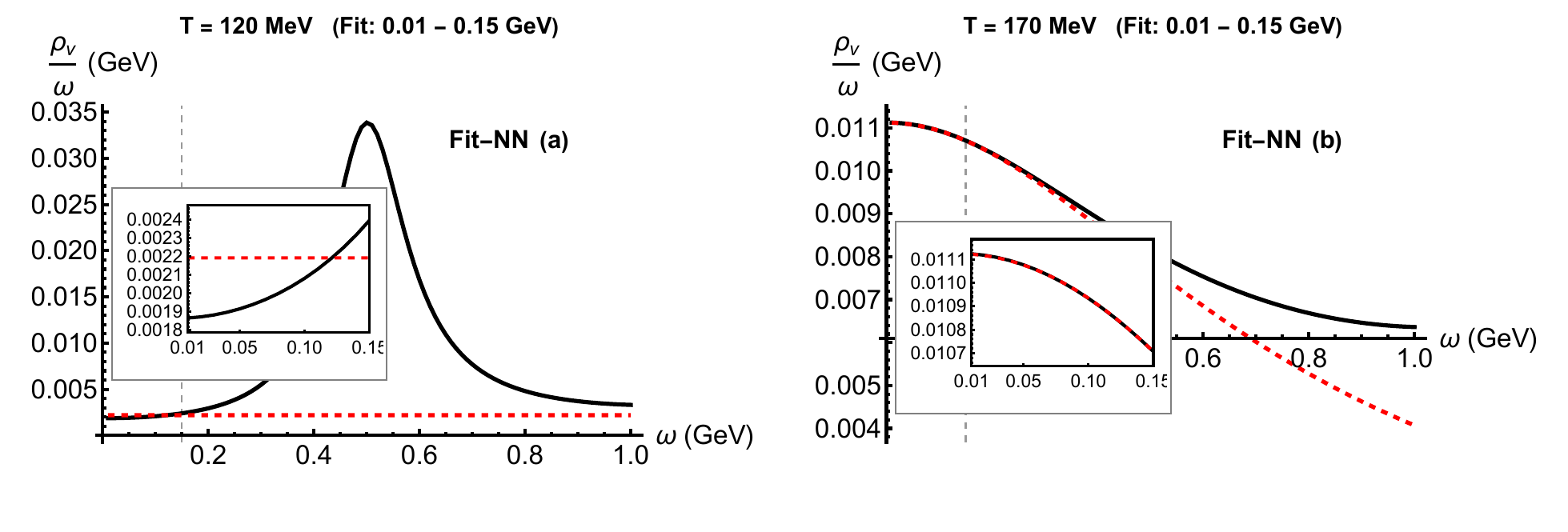}
  \caption{Low-frequency spectral response $\rho_v(\omega)/\omega$ of the transverse vector
  correlator at (a) $T=120~\mathrm{MeV}$ and (b) $T=170~\mathrm{MeV}$, computed using the
  NN-calibrated dilaton field $\Phi(z)$. The black solid curves represent the numerical
  results, while the red dashed curves show fits to the transport-peak form
  $\rho_v(\omega)=A\omega/(\omega^2+\omega_{\mathrm{Im}}^2)$ over the frequency range
  $0.01\leq\omega\leq0.15~\mathrm{GeV}$.
  The fitted parameters are $A=2.4762~\mathrm{GeV}^{3}$ and
  $\omega_{\mathrm{Im}}^{\mathrm{fit}}=33.5993~\mathrm{GeV}$
  at $T=120~\mathrm{MeV}$, and $A=0.006422~\mathrm{GeV}^{3}$ and
  $\omega_{\mathrm{Im}}^{\mathrm{fit}}=0.75984~\mathrm{GeV}$
  at $T=170~\mathrm{MeV}$.
  The insets provide enlarged views of the fitting region.}
  \label{fig:SFRP}
\end{minipage}
\end{figure}

The computed spectral functions are summarized in Figure~\ref{fig:SFNN}, which shows results for $T=100$--$160~\mathrm{MeV}$. At $T=100~\mathrm{MeV}$ [Figure~\ref{fig:SFNN}(a)], Fit-NN yields $m_{\mathrm{pole}}=0.58~\mathrm{GeV}$ and $\Gamma/2=0.002~\mathrm{GeV}$, giving a peak at $\omega_{\mathrm{peak}}=0.58~\mathrm{GeV}$. Since $m_{\mathrm{pole}}\gg \Gamma/2$, the spectrum exhibits a narrow Breit-Wigner resonance whose peak position closely agrees with $m_{\mathrm{pole}}$. At $T=120~\mathrm{MeV}$ [Figure~\ref{fig:SFNN}(b)], the peak broadens as the width increases, and Fit-NN gives $m_{\mathrm{pole}}=0.50~\mathrm{GeV}$ and $\Gamma/2=0.09~\mathrm{GeV}$ with $\omega_{\mathrm{peak}}=0.51~\mathrm{GeV}$. Once $\Gamma/2$ becomes non-negligible, the resonance is no longer governed solely by $m_{\mathrm{pole}}$; both the peak height and the peak position are jointly influenced by $m_{\mathrm{pole}}$ and the thermal width $\Gamma$, so that $\omega_{\mathrm{peak}}$ need not coincide with $m_{\mathrm{pole}}$. At $T=140~\mathrm{MeV}$ [Figure~\ref{fig:SFNN}(b)], Fit-NN yields $m_{\mathrm{pole}}=0.44~\mathrm{GeV}$ and $\Gamma/2=0.31~\mathrm{GeV}$. In this regime, the half-width $\Gamma/2$ is comparable to $m_{\mathrm{pole}}$, while the full thermal width $\Gamma$ already exceeds $m_{\mathrm{pole}}$. The resonance becomes strongly broadened, making it difficult to identify a well-defined quasiparticle peak. Consequently, the local maximum of the spectrum can deviate markedly from $m_{\mathrm{pole}}$, and we find $\omega_{\mathrm{peak}}=0.57~\mathrm{GeV}$ despite $m_{\mathrm{pole}}=0.44~\mathrm{GeV}$. Concurrently, the low-frequency (transport) weight increases, shifting part of the overall spectral strength toward small $\omega$.
To further examine this evolution, Figure~\ref{fig:SFRP} compares $\rho_v(\omega)/\omega$ at $T=120$ and $170~\mathrm{MeV}$ using the transport-peak form $\rho_v(\omega)=A\omega/(\omega^2+\omega_{\mathrm{Im}}^2)$, with both $A$ and $\omega_{\mathrm{Im}}$ treated as fitting parameters. At $T=120~\mathrm{MeV}$, the low-frequency behavior cannot be reproduced by this form, since $\rho_v(\omega)/\omega$ increases with $\omega$, whereas the transport-peak form predicts a monotonically decreasing behavior of $\rho_v(\omega)/\omega$. In contrast, at $T=170~\mathrm{MeV}$, the numerical result is well described by the same form in the low-frequency region, supporting the emergence of a transport peak. This confirms the near-pole behavior predicted by the QNM analysis, showing that the spectral function indeed evolves toward a transport peak as the real part of the QNM frequency vanishes. At higher frequencies, the deviations become more pronounced due to additional frequency-dependent contributions to the full holographic spectral function that are not captured by the simple low-frequency approximation.

\onecolumngrid
\twocolumngrid

\onecolumngrid
\twocolumngrid

Combining the spectral functions in Figure~\ref{fig:SFNN}, the low-frequency analysis in Figure~\ref{fig:SFRP}, the quasinormal-mode results in Figure~\ref{fig:M-NN-mpg}, and the peak analyses at $T=100$, $120$, and $140~\mathrm{MeV}$, we obtain a consistent picture. At low temperature, where $\Gamma/2 \ll m_{\mathrm{pole}}$, the spectrum shows a narrow Breit--Wigner-like resonance whose position closely tracks $m_{\mathrm{pole}}$. With increasing temperature, $m_{\mathrm{pole}}$ decreases while $\Gamma/2$ grows; the peak broadens and the position $\omega_{\mathrm{peak}}$ no longer coincides with $m_{\mathrm{pole}}$, being jointly influenced by both $m_{\mathrm{pole}}$ and $\Gamma/2$, which signals the breakdown of a purely single-quasiparticle description. A practical indicator of the onset of a strongly damped regime in our setup, as shown in Figure~\ref{fig:M-NN-mpg}, is the crossing $\Gamma/2=m_{\mathrm{pole}}$ at $T\simeq147~\mathrm{MeV}$. As the temperature increases further, the resonance becomes progressively less well defined, while the spectral response evolves toward a broad, transport-dominated low-frequency structure. At still higher temperatures, where $\Gamma/2 \gtrsim m_{\mathrm{pole}}$, the spectrum becomes strongly broadened with enhanced low-frequency (transport) weight; any residual local maximum can substantially deviate from $m_{\mathrm{pole}}$. In short, as $T$ increases the line shape evolves from a narrow Breit--Wigner resonance ($\Gamma/2 \ll m_{\mathrm{pole}}$), to a broadened structure where $\omega_{\mathrm{peak}}$ departs from $m_{\mathrm{pole}}$, and eventually toward a transport-weighted spectrum, with this evolution occurring gradually rather than at a sharply defined temperature.

\subsection{Dispersion Relation and Thermal Propagator}\label{subsec:dispersion}

To gain a more complete picture of the in-medium dynamics, we study the dispersion relation of the vector mode. For a transverse plane-wave ansatz $v_m(x,z)=\varepsilon_m\,v(\omega,p;z)\,e^{-i\omega t+i\vec p\cdot\vec x}$ with $\vec p\!\cdot\!\vec\varepsilon=0$, the equation of motion reads
\begin{equation}
\begin{gathered}
v''(z)+\Bigl(A'(z)-\Phi'(z)+\frac{f'(z)}{f(z)}\Bigr)\,v'(z)\\
+\Bigl(\frac{\omega^{2}}{f(z)^{2}}-\frac{p^{2}}{f(z)}\Bigr)\,v(z)=0.
\end{gathered}
\label{eq:drtp}
\end{equation}
Equation~\eqref{eq:drtp} reduces to the static Eq.~\eqref{eq:v-ODE} for $\omega=0$, and to the zero-spatial-momentum Eq.~\eqref{eq:v-ODEmg} for $p=0$. Here, $\omega$ is the complex frequency, and $p=|\vec p|$ is the spatial momentum. The near-boundary expansion is given by
\begin{equation}
v(z \to 0) = v_0 + v_2 z^2 + \frac{1}{2} v_0 (p^2-\omega^2) z^2 \log(z) + \mathcal{O}(z^3),
\label{eq:drtpz0}
\end{equation}
while at the horizon we impose the ingoing condition,
\begin{equation}
v(z \to z_h) = (z_h - z)^{-\frac{i \omega}{4 \pi T}}
\Bigl[\,v_{h0} + v_{h1}\,(z_h - z) + \mathcal{O}\bigl((z_h - z)^2\bigr)\Bigr] ,
\label{eq:drtpzh}
\end{equation}
with
\begin{equation}
v_{h1}
=\frac{\bigl(2\omega + 4\,\mathrm{i}\,p^{2} z_h
 - 3\,\mathrm{i}\,\omega^{2} z_h - 4\,\omega z_h\,\Phi'(z_h)\bigr)}
 {8\,(2\mathrm{i}+\omega z_h)}\,v_{h0}\,.
\end{equation}

Based on Eq.~\eqref{eq:drtp} and the boundary conditions Eqs.~\eqref{eq:drtpz0} and \eqref{eq:drtpzh}, we compute the dispersion relation $\omega(p)$ by integrating from the near-horizon region toward the UV boundary, imposing the source-free condition $v_0(\omega,p)=0$. For $p^2<0$, we use the analytic continuation $ip$. 

Guided by numerical data, we extend the zero-width dispersion relation
\begin{equation}
\omega(p)^2 = u^2\vec{p}^{\,2} + m^2,
\label{eq:wp0}
\end{equation}
to a finite temperature form including thermal damping
\begin{equation}
\omega(p) = -i(a + D \vec{p}^2) \pm \sqrt{u^2\vec{p}^{\,2} + m^2},
\label{eq:wp}
\end{equation}
where $a$ characterizes the momentum-independent damping, $D$ is the diffusion coefficient, $u$ is the effective propagation speed\footnote{The parameter $u$ is inherited from the zero-width relation $\omega^2 = u^2 p^2 + m^2$ and is not the strict group or phase velocity. For a massive mode, $v_g = u^2 p/\omega$ and $v_p = \omega/p$, which approach $u$ only asymptotically as $p \to \infty$. A finite width further weakens the quasiparticle interpretation, but $u$ remains a velocity-related parameter; hence we simply call it an effective speed.}, and $m^2$ is a parameter of the dispersion relation. These parameters are determined by fitting the numerical quasi-normal mode dispersion relation.
This dispersion relation yields the finite temperature propagator
\begin{equation}
    G(\omega,\vec{p}) \sim
    \frac{1}{
      \big[\,\omega + i\big(a + D\,\vec{p}^{\,2}\big)\,\big]^2
      - \big(u^{2}\vec{p}^{\,2} + m^{2}\big)
    }.
\label{eq:prop}
\end{equation}

\begin{figure}[H]
\centering
\begin{minipage}{1\textwidth}
  \centering
  \includegraphics[width=\linewidth]{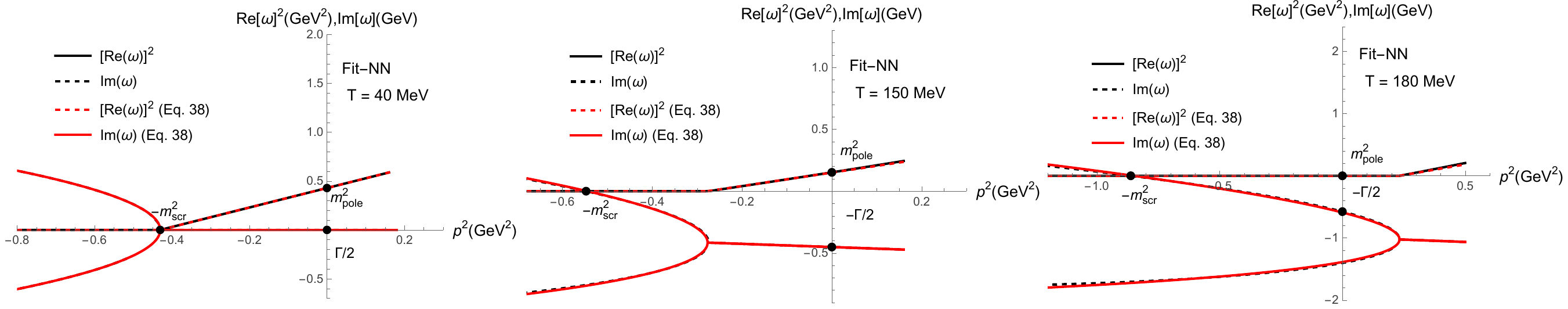}
  \caption{Dispersion relation at three temperatures, $T=40$ (left panel), $150$ (middle panel), and $180~\mathrm{MeV}$ (right panel).
           Black curves, direct numerical solutions of Eq.~\eqref{eq:drtp};
           red curves, fits to the finite-temperature ansatz in Eq.~\eqref{eq:wp}.
           Each panel displays the curves $[\mathrm{Re}\,\omega(p)]^2$ vs.\ $p^{2}$ and $\mathrm{Im}\,\omega(p)$ vs.\ $p^{2}$.
       The fitted parameters of Eq.~\eqref{eq:wp} are
$a=0$, $D=0$, $u^{2}=0.9961$, $m^{2}=0.4290~\mathrm{GeV}^{2}$ at $T=40~\mathrm{MeV}$;
$a=0.4513~\mathrm{GeV}$, $D=0.1292~\mathrm{GeV}^{-1}$, $u^{2}=0.5425$, $m^{2}=0.1501~\mathrm{GeV}^{2}$ at $T=150~\mathrm{MeV}$;
and $a=0.9887~\mathrm{GeV}$, $D=0.1508~\mathrm{GeV}^{-1}$, $u^{2}=0.6837$, $m^{2}=-0.1591~\mathrm{GeV}^{2}$ at $T=180~\mathrm{MeV}$.}
  \label{fig:NNDRTPP}
\end{minipage}
\end{figure}

As shown in Figure~\ref{fig:NNDRTPP}, at three temperatures,
$T=40$, $150$, and $180~\mathrm{MeV}$, we present
$[\mathrm{Re}\,\omega(p)]^2$ versus $p^2$
(black solid curves calculated from Eq.~\eqref{eq:drtp}
and red dashed curves fitted using Eq.~\eqref{eq:wp}),
and $\mathrm{Im}\,\omega(p)$ versus $p^2$
(black dashed curves calculated from Eq.~\eqref{eq:drtp}
and red solid curves fitted using Eq.~\eqref{eq:wp}).
Since Eq.~\eqref{eq:wp} is a fitted parameterization, small differences between the fitted intercepts and the corresponding numerical results may arise from fitting and rounding errors.

\onecolumngrid
\twocolumngrid

At $T=40\mathrm{MeV}$, for the Fit-NN model, the quasi-normal modes analysis gives
$m_{\mathrm{pole}}^{2}=0.43~\mathrm{GeV}^{2}$, $\Gamma/2=0$, and
the static screening point $-m_{\mathrm{scr}}^{2}=-0.43~\mathrm{GeV}^{2}$.
In this regime the extended ansatz Eq.~\eqref{eq:wp} reduces to its zero-width form Eq.~\eqref{eq:wp0};
the real branch is a straight line
$[\mathrm{Re}\,\omega(p)]^{2}=u^{2}p^{2}+m^{2}$ with $u^{2}=0.9961$,
while the imaginary branch remains zero for $p^{2}\ge -0.43~\mathrm{GeV}^{2}$ (fitted $D=0$).
Consequently, the $x$-intercept of the real branch approximately coincides with the true static point, and the intuitive relation
$-m_{\mathrm{scr}}^{2}\simeq-m_{\mathrm{pole}}^{2}/u^{2}$
holds within the numerical fitting accuracy.

At $T=150\mathrm{MeV}$, for the Fit-NN model, the quasi-normal modes analysis yields
$m_{\mathrm{pole}}^{2}=0.15~\mathrm{GeV}^{2}$, $\Gamma/2=0.45~\mathrm{GeV}$, and a screening mass satisfying
$-m_{\mathrm{scr}}^{2}=-0.55~\mathrm{GeV}^{2}$;
solving the static condition $\omega=0$ in Eq.~\eqref{eq:drtp} gives the same $-m_{\mathrm{scr}}^{2}=-0.55~\mathrm{GeV}^{2}$.
If one neglects the width and uses the zero-width relation Eq.~\eqref{eq:wp0}, the real-branch $x$-intercept appears near
$-0.27~\mathrm{GeV}^{2}$, to the right of the true static point.
This displacement highlights a crucial role of the finite width in the static condition $\omega=0$: for sufficiently negative $p^2$ ($p^2<-m^2/u^2$), the square-root term $\sqrt{u^2 p^2 + m^2}$ itself becomes imaginary and contributes to the total imaginary part of $\omega$.
The correct static point is therefore recovered only when this contribution cancels against the momentum-dependent damping term $-i(a + D p^2)$, which is precisely what the full finite-temperature ansatz Eq.~\eqref{eq:wp} captures.

At $T=180\mathrm{MeV}$, according to quasi-normal modes, the pole mass has vanished ($m_{\mathrm{pole}}^{2}=0$), while the screening mass and width remain finite.
For Fit-NN we find $\Gamma/2=0.58~\mathrm{GeV}$ and $-m_{\mathrm{scr}}^{2}=-0.86~\mathrm{GeV}^{2}$.
In this regime, the fitted parameter \(m^2\) becomes negative, so the zero-width relation Eq.~\eqref{eq:wp0} no longer admits the usual quasiparticle interpretation and would yield a positive \(x\)-intercept. The finite-temperature form Eq.~\eqref{eq:wp}, however, continues to reproduce the full numerical solution.
The imaginary branch intersects the $y$-axis at $p=0$ at $-\Gamma/2$ and the $x$-axis at $-m_{\mathrm{scr}}^{2}$.

\begin{figure}[H]
  \centering
  \includegraphics[width=0.4\textwidth]{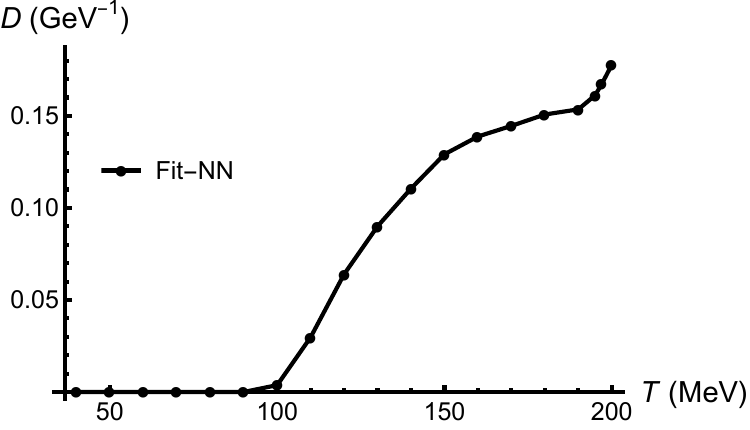}
  \caption{Temperature dependence of the diffusion coefficient $D$. The dots are the calculated data points, and the solid line is a simple interpolation between them.}
  \label{fig:TD}
\end{figure}

\begin{figure}[H]
  \centering
  \includegraphics[width=0.4\textwidth]{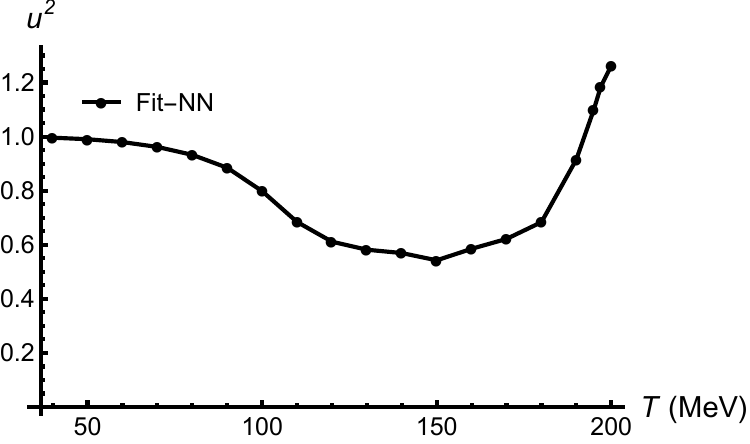}
  \caption{Temperature dependence of the effective propagation speed squared $u^2$. The dots denote the numerically calculated values, and the solid line is a simple interpolation through them.}
  \label{fig:TC}
\end{figure}

This evolution is further corroborated by the temperature dependence of the transport coefficients, as shown in Figures~\ref{fig:TD} and \ref{fig:TC}. The diffusion coefficient $D(T)$ vanishes at low $T$ and becomes nonzero precisely when the quasi-normal mode analysis first yields a finite width, growing monotonically with temperature thereafter. Meanwhile, the effective propagation speed squared $u^2(T)$ remains close to unity in the low-$T$ regime, where the zero-width relation suffices. Once a finite width develops, $u^2$ decreases, and the full form Eq.~\eqref{eq:wp} is required. At higher $T$, $u^2$ turns upward and eventually exceeds unity, at which point the parameterization ceases to admit a simple quasiparticle-velocity interpretation, signaling the breakdown of a particle-like description.

In summary, the dispersion relation evolves from a zero-width quasiparticle form at low $T$, to a finite-width regime where the static point is shifted by the imaginary part of the dispersion relation, and finally to a high-$T$ regime where the quasiparticle picture breaks down. While the zero-width relation Eq.~\eqref{eq:wp0} fails once the width becomes sizable, the full finite-temperature ansatz Eq.~\eqref{eq:wp} remains valid and reproduces the numerical solutions of Eq.~\eqref{eq:drtp} across all temperatures.

\section{Conclusion and discussion}\label{sec:conclusion}

In this work, we employ a finite-temperature soft-wall AdS/QCD model and use machine learning to calibrate the dilaton field. With the calibrated dilaton field, we reproduce the lattice screening masses in quantitative agreement with lattice QCD. On this background, we compute the quasi-normal modes, the spectral functions, and the dispersion relation of the $\rho$ meson.

From the quasi-normal mode analysis, we find that $m_{\mathrm{pole}}$ stays close to $m_{\mathrm{scr}}$ at low temperatures, decreases as the temperature rises, and vanishes at $T_d\approx168~\mathrm{MeV}$, indicating that the narrow Breit-Wigner-like resonance peak evolves into a transport peak. This picture is qualitatively consistent with the lattice findings of Ref.~\cite{Brandt:2015aqk}, where the quasiparticle peak was found to melt and give way to a substantial low-frequency spectral weight in the same temperature regime. Motivated by this consistency, we adopt the vanishing real part of the quasi-normal mode frequency, \(\omega_{\mathrm{Re}}(T_d)=0\), as an operational dissociation criterion.

The spectral functions provide a detailed picture of this evolution. In the low-temperature regime, the width is numerically negligible and the spectrum exhibits a narrow Breit--Wigner-like resonance whose peak position $\omega_{\mathrm{peak}}$ closely coincides with $m_{\mathrm{pole}}$. As the temperature increases to the intermediate regime, the peak broadens and $\omega_{\mathrm{peak}}$ starts to deviate from $m_{\mathrm{pole}}$, since both the peak height and position are jointly influenced by $m_{\mathrm{pole}}$ and the thermal width $\Gamma$. In the high-temperature regime, the half-width becomes comparable to the pole mass, the resonance is strongly broadened, and the local maximum of the spectrum can deviate markedly from $m_{\mathrm{pole}}$. Concurrently, the low-frequency (transport) weight increases, shifting part of the overall spectral strength toward small $\omega$. We also examine the low-frequency region of the spectral functions at high temperature, and the results confirm the QNM prediction that the vanishing of the pole mass is accompanied by the emergence of a transport peak, which is consistent with the dissociation picture of Ref.~\cite{Brandt:2015aqk}.

Complementing the quasi-normal mode analysis and the spectral functions, the dispersion relation offers a momentum-resolved picture of the same in-medium dynamics. It is consistent with the quasi-normal mode analysis: the vanishing of $m_{\mathrm{pole}}$ corresponds to the $[\mathrm{Re}\,\omega(p)]^2$ branch passing through the origin at $p^2=0$. More importantly, it reveals the critical role of finite width in the static condition. At low $T$, where the width is negligible, the static screening point coincides with the zero-width $x$-intercept and the dispersion reduces to Eq.~\eqref{eq:wp0}. As temperature increases, the emerging finite width shifts the true static point leftward relative to the naive zero-width intercept. This shift originates from the imaginary contribution of the square-root term at negative $p^2$, which cancels against the momentum-dependent damping term $-i(a + D p^2)$. The full finite-temperature ansatz Eq.~\eqref{eq:wp} correctly captures this mechanism and reproduces the numerical solutions of Eq.~\eqref{eq:drtp} across all temperatures. At high $T$, where $m_{\mathrm{pole}}\to 0$ while $\Gamma/2$ and $m_{\mathrm{scr}}$ remain finite, the zero-width relation fails entirely, whereas the finite-temperature form continues to match the full solutions. Guided by this dispersion relation, we refine the finite-temperature propagator to
\[
G(\omega,\vec{p}) \sim
    \frac{1}{
      \big[\,\omega + i\big(a + D\,\vec{p}^{\,2}\big)\,\big]^2
      - \big(u^{2}\vec{p}^{\,2} + m^{2}\big)
    } \, .
\]
Within this framework, the dispersion relation modifies the connection between the screening mass and the pole mass by incorporating finite-temperature damping. This suggests that for a hadron gas, the simple zero-width pole approximation may break down well before the phase transition, and the corrections encoded in $G(\omega,\vec{p})$ can become significant already at temperatures below $T_d$. Such finite-width effects should therefore be taken into account in future phenomenological studies of hadronic matter.

We also examined the temperature dependence of the transport coefficients. The diffusion coefficient $D(T)$ vanishes at low $T$ and becomes nonzero precisely when the quasi-normal mode analysis first yields a finite width, growing monotonically thereafter. The effective propagation speed squared $u^2(T)$ remains close to unity in the low-$T$ regime, decreases once a finite width develops, and eventually turns upward, exceeding unity at higher $T$. When $u^2>1$, the parameterization ceases to admit a simple quasiparticle-velocity interpretation, signaling the breakdown of a particle-like description.

With the background geometry fixed, the vector-channel properties in the soft-wall model can be used to constrain the dilaton field $\Phi$ using lattice data. On this basis, one can subsequently include the chiral condensate and the associated $\chi$ field step by step, and ultimately fix the entire model with lattice data so as to make further predictions. We hope that this work will provide a useful reference for future studies.

\vspace*{1cm}

{\bf Acknowledgements}\quad
This work is mainly supported by the National Natural Science Foundation of China under Grant Nos. 12275108, 12235016.
The work of L. Y. is supported by the
NSFC under Grant No. 11605072 and the Seeds Funding of Jilin University.

\appendix
\section*{Appendix}

In the main text, we have restricted our analysis to the spatially transverse components $V_{1,2}$ of the five-dimensional vector field, which are perpendicular to the spatial momentum $\vec p$. Their equations of motion in the static limit, at zero spatial momentum, and in the general case are given by Eqs.~\eqref{eq:v-ODE}, \eqref{eq:v-ODEmg}, and~\eqref{eq:drtp}, respectively.

To complete the description of the vector-field dynamics, in this appendix we extend the analysis to the temporal component $V_0$. Since $V_0$ couples only to the spatial component $V_3$ parallel to $\vec p$, these two components together form the longitudinal channel. As we will show, in the zero-spatial-momentum limit ($p=0$, $\omega\neq0$), $V_0$ does not support an independent dynamical mode, and both the longitudinal and transverse spatial equations, Eqs.~\eqref{eq:app-eomV3} and~\eqref{eq:app_transverse}, reduce to Eq.~\eqref{eq:v-ODEmg} of the main text. Consequently, the pole masses, thermal widths, and operational dissociation temperature obtained in the main text remain unchanged.

The appendix is organized as follows. In Appendix~\ref{app:vector_fields}, we present the complete vector-field action and the corresponding equations of motion. In Appendix~\ref{app:temporal_eom}, we discuss the longitudinal screening mass extracted from the temporal component $V_0$ in the static limit. In Appendix~\ref{app:coupled_eom}, we consider general $(\omega,\vec p)$ and show how $V_0$ and $V_3$ are coupled and jointly determine the dynamics of the longitudinal channel.

Throughout this appendix, $M,N$ denote five-dimensional indices, $\mu,\nu,\rho,\sigma$ denote four-dimensional indices, $m,n=1,2,3$ denote spatial directions, and $i=1,2,3$ denotes the isospin index.

\section{Complete Vector-Field Action and Equations of Motion}
\label{app:vector_fields}

We begin by writing down the complete action for the five-dimensional vector field,
\begin{align}
  S_V&=-\frac{1}{2g_5^2}\!\int\! d^5x\,\sqrt{g}\,e^{-\Phi}\sum_{i=1}^{3}
  \Big\{g^{zz}g^{\mu\nu}\,\partial_zV^i_\mu\,\partial_zV^i_\nu
  \notag\\[-2pt]
  &\qquad\qquad\quad
  +\frac12\,g^{\mu\rho}g^{\nu\sigma}F^i_{\mu\nu}F^i_{\rho\sigma}\Big\}.
  \label{eq:app-SVfull}
\end{align}
For the transverse polarizations, the second term simplifies to
$g^{\mu\nu}g^{mn}\partial_\mu V_m\partial_\nu V_n$, and Eq.~\eqref{eq:app-SVfull} reduces to Eq.~\eqref{eq:Sv} of the main text. For the longitudinal channel, however, the cross terms contained in $F_{\mu\nu}F^{\mu\nu}$ must be retained, since they encode the coupling between the temporal and longitudinal spatial components.

To make this decomposition explicit, we expand the action and obtain
\begin{equation}
\begin{aligned}
S_V^{(2)}
=
\frac{1}{2g_5^2}
\int d^5x\,e^{A-\Phi}
\Bigg\{
&
(\partial_zV_0)^2
-
f\sum_{m=1}^{3}(\partial_zV_m)^2
\\
&+
\frac{1}{f}
\sum_{m=1}^{3}
(\partial_0V_m-\partial_mV_0)^2
\\
&-
\frac{1}{2}
\sum_{m,n=1}^{3}
(\partial_mV_n-\partial_nV_m)^2
\Bigg\},
\end{aligned}
\label{eq:app-expanded}
\end{equation}
where the isospin index has been suppressed. We choose the four-momentum as $q^\mu=(\omega,0,0,p)$ and introduce the Fourier decomposition
\begin{equation}
V_\mu(x,z)
=
V_\mu(z)e^{-i\omega t+ipx^3}.
\end{equation}
The action then separates into two mutually decoupled sectors,
\begin{align}
S_V^{\rm T}
&=-\frac{1}{2g_5^2}
\int d^5x\,e^{A-\Phi}
\sum_{\alpha=1,2}
\Bigl[
f(\partial_zV_\alpha)^2
\notag\\[-2pt]
&\qquad\qquad
-\frac{1}{f}(\partial_0V_\alpha)^2
+(\partial_3V_\alpha)^2
\Bigr],
\label{eq:app-ST}
\\[2pt]
S_V^{\rm L}
&=-\frac{1}{2g_5^2}
\int d^5x\,e^{A-\Phi}
\Bigl[
-(\partial_zV_0)^2
+f(\partial_zV_3)^2
\notag\\[-2pt]
&\qquad\qquad
-\frac{1}{f}
(\partial_0V_3-\partial_3V_0)^2
\Bigr],
\label{eq:app-SL}
\end{align}
where $\alpha=1,2$ labels the transverse spatial directions. The transverse sector $S_V^{\rm T}$ reproduces the dynamics studied in the main text, while the longitudinal sector $S_V^{\rm L}$ contains the coupling between $V_0$ and $V_3$.

Varying the transverse action~\eqref{eq:app-ST} with respect to $V_\alpha$, we obtain
\begin{equation}
\begin{aligned}
V_\alpha''
+
\left(
A'-\Phi'+\frac{f'}{f}
\right)V_\alpha'
+
\left(
\frac{\omega^2}{f^2}
-\frac{p^2}{f}
\right)V_\alpha=0,
\end{aligned}
\label{eq:app_transverse}
\end{equation}
which is identical to Eq.~\eqref{eq:drtp} of the main text. Varying the longitudinal action~\eqref{eq:app-SL} with respect to $V_0$ and $V_3$, respectively, we obtain the coupled equations
\begin{align}
&V_0''
+\bigl(A'-\Phi'\bigr)V_0'
-\frac{p}{f}
\bigl(pV_0+\omega V_3\bigr)=0,
\label{eq:app-eomV0}
\\[2pt]
&V_3''
+\left(A'-\Phi'+\frac{f'}{f}\right)V_3'
+\frac{\omega}{f^2}
\bigl(\omega V_3+pV_0\bigr)=0.
\label{eq:app-eomV3}
\end{align}
In addition, since we have imposed the radial gauge $V_z=0$, the variation with respect to $V_z$ no longer yields an independent equation of motion, but it must be retained as a constraint. From $\partial_M(\sqrt{g}\,e^{-\Phi}F^{Mz})=0$, we find
\begin{equation}
\omega V_0'+pfV_3'=0.
\label{eq:app-constraint}
\end{equation}
Equations~\eqref{eq:app-eomV0}--\eqref{eq:app-constraint} are consistent with Eqs.~(4.4)--(4.6) of Ref.~\cite{Ballon-Bayona:2024twa}.

\section{Equation of Motion for the Temporal Component $V_0$}
\label{app:temporal_eom}

Having established the full set of equations, we now examine two special limits that connect the appendix to the main text. First, in the zero-spatial-momentum limit ($\vec p=0$, $\omega\neq0$), the constraint equation~\eqref{eq:app-constraint} gives $V_0'=0$, so $V_0$ retains only its boundary value and does not support an independent normalizable dynamical mode. Meanwhile, Eq.~\eqref{eq:app-eomV3} reduces to Eq.~\eqref{eq:v-ODEmg} of the main text. This confirms that the pole masses, thermal widths, and dissociation temperature obtained in the main text remain unchanged.

We then turn to the static limit $\omega=0$. Here the constraint equation~\eqref{eq:app-constraint} gives $V_3'=0$, so $V_3$ decouples, and Eq.~\eqref{eq:app-eomV0} reduces to
\begin{equation}
  V_0''+\bigl(A'-\Phi'\bigr)V_0'
  -\frac{p^2}{f}\,V_0=0 .
  \label{eq:app-V0static}
\end{equation}
Following the definition introduced in \autoref{subsec:ml} of the main text, we define the longitudinal screening mass extracted from the static temporal component $V_0$ through the eigenvalue condition $m_{\rm scr,L}^{2}=-p^2$, and denote the transverse screening mass obtained in the main text by $m_{\rm scr,T}\equiv m_{\rm scr}$. It should be emphasized that $m_{\rm scr,L}$ characterizes the inverse spatial correlation length associated with the static charge-density correlator $\langle J_0J_0\rangle(\omega=0,\vec p)$, rather than the mass of an independently propagating temporally polarized $\rho$ meson~\cite{Brandt:2013faa,Cui:2011wb}.

The near-boundary expansions of $V_0$ in the UV and IR regions are given by
\begin{equation}
  V_0(z \to 0) = v_{0} + v_{2}\, z^{2}
  + \frac{1}{2}\, v_{0} p^{2} z^{2}\log(z) + \mathcal{O}(z^{3}) ,
  \label{eq:app-V0UV}
\end{equation}
\begin{equation}
\begin{gathered}
  V_0(z \to z_h) = v_{h1}\,(z_h-z)\\
  + \frac{v_{h1}\left(-4 + p^{2} z_h^{2}
    - 4 z_h \Phi'(z_h)\right)}{8 z_h}\,(z_h-z)^{2}
  + \mathcal{O}[(z_h-z)^{3}] ,
  \label{eq:app-V0IR}
\end{gathered}
\end{equation}
where we set $v_{h1}=1$.

\begin{figure}[H]
  \centering
  \includegraphics[width=0.4\textwidth]{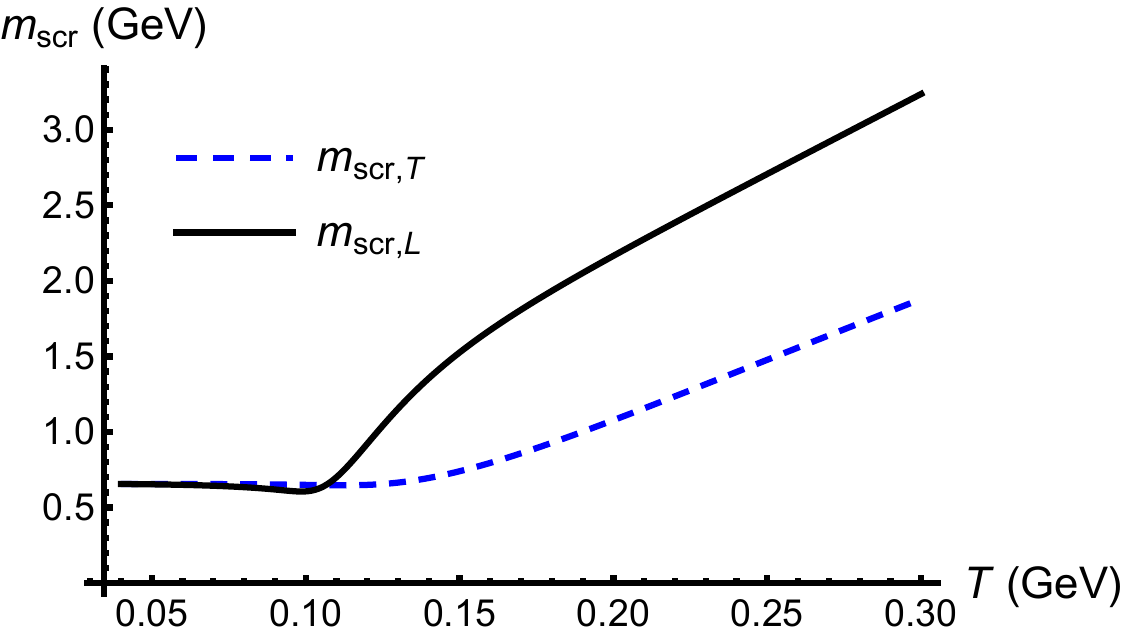}
  \caption{Transverse ($m_{\mathrm{scr},T}$, blue dashed) and longitudinal
  ($m_{\mathrm{scr},L}$, black solid) screening masses as functions of temperature.}
  \label{fig:V0scr}
\end{figure}

The calculated longitudinal and transverse screening masses are presented in Fig.~\ref{fig:V0scr}. As $T\to0$, the blackening function approaches $f\to1$, and Eq.~\eqref{eq:app-V0static} coincides with Eq.~\eqref{eq:v-ODE}. Consequently, in the vacuum limit, $m_{\rm scr,L}=m_{\rm scr,T}$. At finite temperature, the longitudinal and transverse screening masses exhibit a splitting~\cite{Gutierrez-Guerrero:2026hhu}. The splitting between longitudinal and transverse vector screening masses has been investigated using continuum Schwinger-function methods~\cite{Chen:2024emt}, and related longitudinal and transverse screening channels have also been studied in lattice QCD~\cite{Cheng:2010fe}.

\section{Coupled Equations of Motion for $V_0$ And $V_3$}
\label{app:coupled_eom}

For general $(\omega,\vec p)$, the temporal component $V_0$ and the longitudinal spatial component $V_3$ remain coupled, as is evident from Eqs.~\eqref{eq:app-eomV0} and~\eqref{eq:app-eomV3}. To simplify this system, we follow Refs.~\cite{Kovtun:2005ev,Mamani:2013ssa,Sheng:2024kgg} and introduce the gauge-invariant longitudinal combination
\begin{equation}
  E_L\equiv\omega V_3+pV_0 .
  \label{eq:app-E}
\end{equation}
Using Eqs.~\eqref{eq:app-eomV0}--\eqref{eq:app-constraint}, the coupled equations can be reduced to a single independent equation of motion,
\begin{equation}
\begin{gathered}
   E_L''+
  \Bigl[
    A'-\Phi'
    +\frac{\omega^2f'}{f\,(\omega^2-p^2f)}
  \Bigr] E_L'
  +\frac{\omega^2-p^2f}{f^2}\, E_L=0 ,
\end{gathered}
  \label{eq:app-master}
\end{equation}
which is consistent with Eq.~(4.9) of Ref.~\cite{Ballon-Bayona:2024twa}. As a consistency check, we note that in the zero-spatial-momentum limit $\vec p=0$, Eq.~\eqref{eq:app-master} reduces to Eq.~\eqref{eq:v-ODEmg}, while in the static limit $\omega=0$ it reduces to Eq.~\eqref{eq:app-V0static} with $E_L=pV_0$.

The UV asymptotic expansion of $E_L$ is
\begin{equation}
  E_L(z \to 0)
  =e_0+e_2 z^2
  +\frac{1}{2}e_0(p^2-\omega^2)z^2\log(z)
  +\mathcal{O}(z^3),
  \label{eq:app-EUV}
\end{equation}
and near the horizon the ingoing solution admits the expansion
\begin{equation}
  E_L(z \to z_h)
  =(z_h-z)^{-\frac{i\omega}{4\pi T}}
  \Bigl[
    e_{h0}+e_{h1}(z_h-z)
    +\mathcal{O}\bigl((z_h-z)^2\bigr)
  \Bigr],
  \label{eq:app-EIR}
\end{equation}
where
\[
\begin{split}
  e_{h1}
  =\frac{1}{8\,\omega\,(2\mathrm{i}+\omega z_h)}
  \Bigl[
    &\omega^{2}\bigl(2-3\,\mathrm{i}\,\omega z_h\bigr)
    +4\,\mathrm{i}\,p^{2}\bigl(4\,\mathrm{i}+\omega z_h\bigr)\\
    &-4\omega^{2}z_h\Phi'(z_h)
  \Bigr]e_{h0}.
\end{split}
  \label{eq:app-eh1}
\]

These equations provide the boundary conditions needed to numerically solve Eq.~\eqref{eq:app-master}. Together with the transverse equation~\eqref{eq:app_transverse} studied in the main text, they complete the description of the vector-field dynamics in both the transverse and longitudinal channels. The longitudinal channel can be analyzed in a way parallel to the transverse one, and its screening mass is discussed in Appendix~\ref{app:temporal_eom} and shown in Fig.~\ref{fig:V0scr}. We leave a detailed study of the longitudinal spectral function and dispersion relation to future work.

\bibliographystyle{apsrev4-1}
\bibliography{refs}
\end{document}